# Simple Surveys:
# Response Retrieval Inspired by Recommendation Systems

Nandana Sengupta
Azim Premji University

Nati Srebro
Toyota Technological Institute of Chicago
University of Chicago

James Evans
University of Chicago

**Abstract**

In the last decade, the use of simple rating and comparison surveys has proliferated on social and digital media platforms to fuel recommendations. These simple surveys and their extrapolation with machine learning algorithms like matrix factorization shed light on user preferences over large and growing pools of items, such as movies, songs and ads. Social scientists have a long history of measuring perceptions, preferences and opinions, often over smaller, discrete item sets with exhaustive rating or ranking surveys. This paper introduces simple surveys for social science application. We ran experiments to compare the predictive accuracy of both individual and aggregate comparative assessments using four types of simple surveys – pairwise comparisons and ratings on 2, 5 and continuous point scales in three distinct contexts – perceived Safety of Google Streetview Images, Likeability of Artwork, and Hilarity of Animal GIFs. Across contexts, we find that continuous scale ratings best predict individual assessments but consume the most time and cognitive effort. Binary choice surveys are quick and perform best to predict aggregate assessments, useful for collective decision tasks, but poorly predict personalized preferences, for which they are currently used by Netflix to recommend movies. Pairwise comparisons, by contrast, perform well to predict personal assessments, but poorly predict aggregate assessments despite being widely used to crowdsource ideas and collective preferences. We also demonstrate how findings from these surveys can be visualized in a low-dimensional space that reveals distinct respondent interpretations of questions asked in each context. We conclude by reflecting on differences between sparse, incomplete 'simple surveys' and their traditional survey counterparts in terms of efficiency, information elicited and settings in which knowing less about more may be critical for social science.

**Introduction**

Social scientists study behavioral and institutional quantities whose measurement is well-defined and subject to limited interpretation, from income, body-mass-index, and fertility rates to years of formal education. Social scientists are equally committed to the measurement of quantities whose meaning is fundamentally subjective, however, such as beliefs, attitudes, opinions, preferences, and intentions. Considerable debate has arisen regarding the best way to measure these delicate, interior quantities, without changing them in the process of extraction and representation. The General Social Survey, a sociological survey of adults from randomly selected U.S. households, and the National Longitudinal Survey of Youth, a longitudinal survey that follows the lives of a sample of American youth born between 1980-84, both prominently collect data on respondent attitudes and preferences. Values measurement has also been at the center of large international research initiatives such as the World Values Survey (Inglehart et al., n.d.) and Eurobarometer (Inglehart and Reif 1991). Two of the most popular elicitation strategies these and related surveys utilize are subjective "ratings" (Thurstone 1927) and "rankings" (Likert 1932) with regard to beliefs, attitudes, opinions and values. Nevertheless, the methodological debate over whether ratings or rankings are most informative remains open.

Most social science studies in this area have dealt with experiments involving a finite set of items exhaustively rated/ranked by respondents (see for instance: Green and Rao 1970; Alwin and Krosnick 1985; Krosnick and Alwin 1988; Peterson and Wilson 1992; Maio et al. 1996; Ovadia 2004; Rossiter 2010, 73–100). Many researchers argue in favor of ratings because they impose less structure on respondents (Munson and McIntyre 1979), while some argue that by forcing respondents to normalize their choices, rankings lead to more consistent results (Kohn

1977), and others find both types of surveys lead to empirically similar orderings (Feather 1973; Rokeach 1973). Limited research in the social science literature has simultaneously compared multiple simple survey designs, such as rating on continuous scales, multi-point scales (e.g., Likert scales), binary choices, and pairwise comparisons (Alwin and Krosnick 1985; Krosnick and Alwin 1988)[1]. Here, we evaluate and compare these elicitation strategies by predicting ratings on most items using those from a few. This strategy is particularly relevant when the list of items to be compared is very large. We also consider how simple survey designs enable qualitative evaluation of the latent space of interpretations underlying respondent's understanding of the survey elicitation (for another approach also see Alwin and Krosnick 1985).

Social scientists and marketing researchers have a long tradition of assessing values measurement. Over the last decade, similar questions have also become relevant for computer scientists. Social media websites like *Facebook* and *Twitter*; digital media platforms including *Netflix*, *Spotify* and *YouTube*; and e-commerce websites such as *Amazon* and *Alibaba* have actively experimented with different rating systems. Data collected from these websites has characteristics that distinguish them from traditional social science surveys. First, the set of items to be rated is extremely large and often growing – leading to necessarily sparse responses. Second, data is not collected via traditional surveys and random samples but via crowdsourcing convenience samples on digital platforms. Third, machine learning algorithms are used to analyze the data and make predictions about missing quantities (e.g., the song you haven't heard; the movie you haven't watched; the product you haven't bought). An entire subfield of machine learning, "Recommendation Systems", focuses on the efficient estimation of user preferences over items.

Netflix has been at the forefront of this research, amplified in prominence by the much-publicized "Netflix Prize", a competition that bequeathed a million dollars to the team and model that most improved prediction of user movie ratings above their own CineMatch algorithm using only previously recorded ratings (Bennet and Lanning 2007; Bell and Koren 2007). Recently, the company moved from eliciting user responses with a 5 star rating system to a binary "Thumbs Up / Thumbs Down" system. According to Netflix spokespersons (Goode 2017), reasons for this switch include not only the lower cognitive load of the new system leading to a sharp increase in ratings, but also an effort to flatten distinctions users make between movies they like for different reasons (e.g., "guilty pleasures" versus "great films") and so improve the overall correlation between ratings and persistent viewing.

Netflix also shifted from representing user recommendations on a five-star scale in favor of a new "percentage match" feature to signal that starred ratings are not averages, but rather personalized recommendations. This realization further incentivizes users to add personal ratings in order to improve recommendations. Sophisticated reasoning and experiment underlie these changes, which tune the survey design to amplify insights most relevant to their purposes—the provision of promising movie, TV and advertisement suggestions.

A distinctive feature of Recommendation Systems is that they are able to predict comparative assessments over large number of items with a small subset of actual user responses to "simple surveys". By simple surveys, we mean questions that elicit quick responses, which can be gathered at very low cost. Simple surveys are typically incomplete, soliciting respondents to provide assessments on a large and sometimes growing number of items. As such, their use has increased alongside advances in machine learning methods that reconstruct sparse matrices, predicting the large and growing number of missing values.

The statistical method used to reconstruct sparse matrices for recommendations is known as matrix factorization. The matrix factorization framework, illustrated in Figure 1, differs from traditional imputation models in social science based on its (1) ability to handle large, very sparse datasets and (2) use of out-of-sample testing or cross-validation for model selection. Sparse matrix reconstruction has become an active area of computer science research and led to successful recommendation engines. Much early work in the area revolved around Principal Components Analysis (PCA) and Singular Value Decomposition (SVD), where all entries in the data matrix are numerical-valued. The literature has now developed to accommodate heterogeneous data types under a unified matrix completion framework (Collins, Dasgupta, and Schapire 2002; Udell et al. 2016; Singh and Gordon 2008), or even sought to complete very large matrices heuristically using neural networks (Pennington, Socher, and Manning 2014; Levy and Goldberg 2014). Recent efforts have also sought to understand and illustrate the statistical properties of these estimators (Udell et al. 2016). We argue that understanding our own social scientific purposes will enable us to better evaluate and tune simple survey designs. Successful prediction of individual comparative assessments can allow social scientists to use fewer questions to elicit more information from respondents, shortening lengthy surveys and program evaluations. Successful prediction of aggregate comparative assessments, by contrast, enables social scientists to efficiently source dominant shared attitudes, preferences and opinions in order to crowdsource a collective decision or produce a broadly useful service. We also explore the potential for these simple surveys to reveal qualitative differences in respondent interpretation of questions posed.

______________________________

Figure 1 about here

______________________________

In this paper we bridge the gap between the social science literature on the measurement of values and computer science techniques used in recommendation systems. We consider four simple surveys—ratings on a 2 point scale (or binary choice questions), 5 point scale, continuous (infinite) scale and pairwise comparisons.[2] We compare these surveys in terms of cognitive load and predictive ability in a variety of contexts. Data is collected using a panel or crowdsourcing survey alongside missing value estimation carried out using recommendation system algorithms in order to quantify how much each marginal question adds to our understanding of the complete set of respondent assessments.[3] We also use these techniques to visually represent the underlying latent space that characterizes responses. To our knowledge this is the first attempt to evaluate survey designs based on their ability to predict comparative assessment under varying degrees of sparseness. It is also among the first attempts to introduce recommendation system ideas and matrix factorization techniques to the social survey literature. The remainder of the paper is organized as follows. The following section describes our study design and data collection. Then we present a brief introduction to matrix factorization methods for prediction and latent space evaluation. Next, we present results and conclude with a discussion of the place of simple surveys in the context of the social sciences.

**Study Design**

We run experiments to measure values in three distinct contexts: (1) the perception of safety regarding Google Streetview panoramas; (2) the likeability or personal appeal associated with works of fine art; and (3) the hilarity or experienced "funny-ness" of animal GIFS—2 to 3 second animal movies. These surveys can be used to evaluate any type of content, from text to scenarios to video. Here we select three visual cases (two static image cases and one short video case) to facilitate scrutiny of the low dimensional representation of responses, which reveals how respondents vary in the way they interpret questions. We also consider these three cases in order to vary the degree of subjectivity involved. Safety perceptions are likely anchored by a shared assessment scheme, while humor evaluations and art preferences may vary more broadly according to personal tastes and experience.

For each context, we select a set of 100 items over which four different sets of simple surveys are compared. First, we consider ratings on a 2 point scale (R2) or binary choice. For example, "Does this street [Streetview panorama] appear safe?" ["Yes" or "No"]. Second, we pose ratings on a 5 point Likert scale (R5). For example, "Rate how much you like this painting [masterpiece image]". ["Don't like it at all" – 1 – to "Really like it" – 5]. Third, we elicit preferences with ratings on a continuous "slider" scale (R100). For example, "Rate how funny this [animal GIF] is". [Slider calibrated between "Not funny at all" to "Extremely funny"]. Note that while the "continuous" scale appears as a continuous slider from the respondents perspective, it is rastered to a discretized scale from 1 to 100, which is what the abbreviation R100 reflects. Finally, we request pairwise comparison (PC). For example, "Which street appears safer?" ["This street appears safer" – first panorama – and "This street appears safer" – second panorama].

Given the large number of items—100 in each context—obtaining a complete ranking is too time-consuming and cognitively demanding to constitute a "simple survey". Consider how ranking 100 items would require sustained focus for respondents to normalize responses regarding a midsize population of numerous items and, as such, are never used in digital comparison tasks at scale. Pairwise comparisons are included instead to facilitate localized rankings (see Rajkumar and Agarwal 2014 for a discussion on conditions under which rank aggregation of pairwise comparisons converge to an optimal ranking). Data for the study is collected via panel design by crowdsourcing with Amazon's Mechanical Turk.

Mechanical Turk panel respondents, often called workers or "Turkers", receive payment by self-selecting to complete short human intelligence tasks. Because our task involved responding to a simple survey, we refer to Turkers as respondents. Unlike social media respondents, who would not typically complete a lengthy survey, our Turker's were incentivized to produce reasonable and consistent responses to all questions. Each was randomly assigned a distinct item and query type. For example, a respondent may be assigned to a survey on the "Visual Perception of Safety" where survey type is R5, involving rating safety on a 5-point Likert scale. Sample queries for each type are presented in Figure 2.

---

Figure 2 about here

---

For rating surveys R2, R5, R100, each respondent answered 80 queries of a fixed type for 80 randomly selected items, then they answered 20 pairwise comparison queries for 20 randomly selected item-pairs. The 80 rating queries were subsequently evaluated by predicting (1) the 20 explicitly compared items and (2) aggregate comparisons that combined all respondent comparisons. In the parlance of machine learning, respondent's explicit comparison functioned as the "ground truth" regarding individual comparative assessments. For pairwise comparison surveys, each respondent answered 100 pairwise comparisons for 100 randomly selected item-pairs. As with the rating surveys, 80 randomly chosen queries were used to estimate the remaining 20 to measure predictive stability.[4] We ensured for a given respondent that none of the $\frac{n(n-1)}{2} = \frac{100 \times 99}{2} = 4950$ unordered pairwise comparisons possibilities were repeated. Different survey types were then compared in terms of response distribution, time to completion, and performance in predicting individual preferences and global rankings.

We collected data from over 600 respondents across different contexts and survey types. After removing respondents who had not completed all 100 survey questions and equalizing the number of respondents within a given context for the 4 survey types, our final data included 46 respondents for each survey type regarding "Safety Perceptions of Streetviews", 49 respondents for each survey about "the Likeability of Artworks", and 50 respondents for each survey with respect to the "Funnyness of Animal GIFS".

**Matrix Factorization for Prediction and Latent Space Evaluation**

Here, we introduce the reader to the matrix factorization framework commonly used in recommendation systems, which we use to evaluate our simple surveys with respect to how their responses predict subsequent personal and aggregate comparative assessments. We also use matrix factorization to enumerate the underlying latent space that reveals distinct dimensions of respondent interpretation. We denote the total number of respondents *m* and total number of items *n*. Data from all respondents corresponding to each rating survey over a given set of items (e.g., R5 on the Safety Perception of Streetviews) can be represented by a matrix *X*, where rows 1 through *m* correspond to respondents and columns 1 through *n* correspond to items that could be rated. Therefore, element $x_{i,j}$ represents respondent *i*'s rating of item *j*. Matrix factorization algorithms fill in elements $x_{i,j}$ for all pairs, even when most matrix entries are missing. Pairwise comparisons lead to related but slightly different specifications and algorithms (See Joachims and Others 2003; Carterette et al. 2008). As the name suggests, matrix factorization techniques involve factoring tabular data into two component matrices *U* and *V* such that when these are multiplied, we recover the original matrix, as illustrated in Figure 1. We detail our specific approach to fit matrix factorization models to our simple survey data in the Appendix.

The *k* columns of matrix *U* represent the latent dimensions for each respondent. Similarly, the *k* rows of matrix *V* represent the latent dimensions for each item. These latent dimensions are ordered in decreasing value by the variance they explain of the original respondent-by-item matrix. Therefore, plotting each respondent according to their values $u_1$ and $u_2$ (the first and second of *k* vectors in *U*) positions them within the two dimensional space in which they vary most. Similarly, plotting items by $v_1$ and $v_2$ (the first and second of *k* vectors associated with *V*) positions them within the 2D space in which they vary most. We utilize this property to array all visual items in the $\{v_1,v_2\}$ latent feature space to reveal respondent interpretations for all three contexts in Figure 7.

**Results**

**Data Summary.** Figure 3 presents the distribution of responses to all queries of type R2, R5 and R100 across all individuals. We note that when given a range of values from which to choose, individuals tend to choose moderate values (the R5 and R100 distributions have less mass at extreme values). This contrasts with R2 surveys where individuals are forced to choose an extreme value (i.e., safe or not safe / like or dislike / funny or not funny). Larger scales allow respondents to hedge their initial uncertainty and calibrate their values on future items. These distributions also confirm that context matters. For example, note that the distribution of R5 in the context of Funny Animal GIFs has a comparatively larger mass at 1 than the other two contexts. This suggests greater *a priori* confidence in the assessment of objects that were experienced as very funny or not at all funny.

______________________________

Figure 3 about here

______________________________

**Comparison of cognitive load.** Figure 4 presents variations in time taken to answer query types R2, R5, R100, and PC as a rough proxy of the cognitive load involved in the distinct queries. Respondents were motivated to answer questions quickly, because completion of our survey would free them to work for compensation on other Turker tasks. We look at the median time taken to answer every set of 8 queries—i.e., the first 8, 9th to 16th, 17th to 24th, ..., and 65th to 72nd queries which correspond to the 10%, 20%, …, and 90% of questions that each respondent answers. We compared medians rather than means because they were less affected by extreme values, to which online surveys may be sensitive. As expected, we find that median times for all tasks across all contexts decreased as respondents answered more questions. Ratings on a continuous slider (R100) consistently took between 40% and 80% more time than all other query types. For example, the median time to respond to the first 8 questions on the continuous slider ranged from 50 to 60 seconds, or from 6 to 8 seconds per question, across all contexts. By contrast,the first 8 questions from other question types took between 30 and 35 seconds, or 3 to 4 seconds per question. The gap reduces with more questions answered subsequently but stays consistently positive. This makes intuitive sense because unlike other query types, R100 required respondent to choose from an infinite number of options, even though the underlying hidden scale was 100 points. This required considerably more calibration and may also have involved greater memory to recall approximate ratings given to previously rated items.

_______________________________

Figure 4 about here

_______________________________

**Predicting Individual Comparative Assessments.** As described above, each respondent received a survey asking 80 questions of a given survey type regarding a particular context. In order to compare each survey type, we used these questions as training data for a matrix factorization model. Each respondent also answered 20 pairwise comparisons, which we used as test data. As an example, consider our prediction of respondent comparative assessmentsbetween Streetviews, using a survey with a 5-point scale and where we predict those comparisons with all 80 of their ratings. With 46 respondents and 100 Streetviews, we have a matrix of dimension 46×100. In this matrix, each respondent has a rating entry (1-5) for 80 of the 100 Streetviews. In order to compare different question types, we randomly sampled 1000 independent draws of 8,

16, 24,..., and 72 responses from each respondent's training data, which corresponded to 10%, 20%,..., and 90% of the total sample. We then used these selective samples to fit a matrix factorization model, which predicted ratings for all omitted items.

In order to measure how surveys and the matrix factorization models built on them performed, we simply summed the number of times the model accurately predicted each of the 20 direct comparisons provided by each individual. The proportion of times predictions were wrong for each individual provides us with an individual test error. The mean of all individual test errors provides us with the model test error. We then compared test errors for each question type and context. For a given number of questions answered, the survey type with the lowest test error is most informative about the remaining, unasked individual comparative assessments.

Figure 5 presents results for predicting individual comparative assessments for each type of simple survey. As expected, with more questions per respondent, we always perform better at predicting individual comparative assessments. We note that even when very few questions are considered, however, such as 8, these techniques are able to beat a random coin toss and sometimes do much better—nearly 20% better than random for continuous scale ratings of animal GIF hilarity. This implies that even if we have very sparse data from each respondent, consolidating data from a moderate number of respondents can be used to make meaningful predictions. Social scientists often ignore such sparse data, and this result suggests that this nevertheless contains reliable information (see Vul et al. 2014 for an extreme example).

When we consider the upper bound of actual comparison prediction, we see that is much better than chance. Our best prediction involves 15% error for an R100 survey regarding the hilarity of animal GIFs when factorization models had been trained on each respondent's answers to all 80 rating questions. Recall that each predicted comparison drew from 100 items. With only 80 items have been seen, this means that a predicted comparison had a 0.36 probability that the respondents had never rated either one or both comparison items before, or 1 minus the probability that they had already rated them both (0.64). In the case that our factorization model was built on only 8 responses regarding humorous animal GIFs, the corresponding probability that they had not rated at least one comparison item was 0.99, but the test error was only 32%. Matrix factorization methods and questions randomized across persons enable these predictions by discovering and exploiting similarities between respondents and similarities between items. As a result, even a very small number of respondent assessments can enable substantial predictive insight about questions they have not been asked.

  In all three contexts, we see consistent patterns. R100 consistently does best at predicting individual comparative assessments. This is followed by R5 and PC, which are similar, and finally R2, which systematically performs worst. From this we infer that giving individuals the ability to calibrate their answers more finely leads to better quality predictions of unasked individual comparisons. Additional degrees of freedom in R5 and R100 enable respondents to calibrate and differentiate their responses to items that lead to information gain. From an information theoretic perspective, it is notable that pairwise comparisons, which contain only 1 bit of information per ranking, rival 5-point Likert scales, which contain up to 5/2 or 2.5 bits per rating. This suggests that pairwise ordinal comparisons compensate for their theoretical information limit by self-calibrating and consistently providing the full bit. Respondents to a five-point rating survey drift, such that responses contain much less information than their theoretical limit.

We note some differences across contexts. First, for funny animal GIFS, we are able to predict individual comparative assessments more accurately than for the likability of artwork or the perceived safety of Streetviews, contrary to our expectations. For animal GIFS even with very sparse data—only 16 questions per respondent—the data from ratings on a continuous scale is able to predict pairwise preferences with a 75% average accuracy. This level of accuracy is not matched for artwork and Streetview panoramas even with 72 questions per respondent. Finally, gains in the prediction of Streetview and animal GIF comparative assessments came quickly then tapered off with more answers considered. By contrast, predictions of artwork likeability linearly increased with more artwork evaluated, suggesting that respondents were discovering rather than revealing their art preferences over the course of the survey. We note anecdotally that survey respondents in the artwork survey, unlike the other surveys, emailed us for references to the artwork they evaluated and expressed gratitude for the opportunity to participate, likely because of the personal discovery and clarification the process enabled.

______________________________

Figure 5 about here

______________________________

**Predicting Aggregate Comparative Assessments.** If we ignore heterogeneous assessments across individuals, we can estimate a global rank order for all entries with relatively few questions by pooling all queries in a given context to estimate an aggregate or global rank order for items. Understanding the aggregate rank order of assessments is useful for any collective decision process, such as an election or poll, which seeks to trigger outcomes desirable for the most respondents. To generate the aggregate rank order for rating surveys (R2, R5 and

R100), we took the mean rating for each item from all users. Sorting these ratings provides us with an estimated global ranking. In the case of R5 and R100, we normalized each individual's rating by subtracting the mean and dividing by the standard deviation across all ratings by that respondent. This ensured that the moments of the distribution for all individual ratings are matched across all individuals. For PC, we applied an aggregation technique known as the borda score (Dwork et al. 2001). For each item pair ($i,j$), we first calculate the proportion $p_{ij}$ of times item $i$ beats $j$. The borda score for $i$ is the mean of $p_{ij}$ for all $j$, or $bs_i = (1/99) \sum_{j \neq i} p_{ij}$. Items are sorted by $bs_i$ to determine aggregate comparative assessment.

To measure the accuracy of each survey type to predict aggregate comparative assessments, we evaluate estimated comparisons against aggregated test data. To generate the aggregate test data, we first built 100 ×100 data matrix that constitutes our aggregate test set. Entries in each row and column correspond to the 100 items ranked. The $(i,j)^{th}$ entry in the matrix is the number of times item $i$ was preferred in a pairwise comparison to item $j$ across respondents. Test error is simply the proportion of times estimated global rankings incorrectly predict pairwise comparisons in the aggregate test data. We estimated global rankings for different sparsity levels by randomly picking 1000 independent draws of 8,16,..., and 72 questions per respondent. We then compared the average global test errors for these subsamples across survey types at different levels of sparsity for each survey context in Figure 6.

As with individual comparisons, here again we see that with more questions per respondent, we can better predict aggregate comparisons, but even with very few questions per respondent, such as 8, these techniques substantially beat a fair coin toss. This implies that even if we have very sparse responses from each respondent, consolidating data from a moderately large number of respondents can produce meaningful aggregate predictions.

R2 performed worst in predicting individual comparisons, but it consistently competes with R100 for best predicting aggregate comparisons. This suggests that the individual variance missed by R2 largely corresponds with the idiosyncratic variance not shared across respondents. Conversely, pairwise comparisons, by efficiently enabling individual calibration, perform worst in every context.

Finally, we note variations in predicting aggregate versus individual comparisons in different contexts. For the likability of masterpieces and the funniness of animal GIFs, we note that even with the largest number of responses per participant, we predict individual comparisons better than aggregate ones. On the other hand, we predict aggregate safety comparisons for Streetviews much more accurately than individual assessments, suggesting far less variance across respondents. Likeable masterpieces and funny animal GIFS are matters of more idiosyncratic taste, and thus less suited to aggregation.

———————————————

Figure 6 about here

———————————————

**Understanding the Latent Space of Responses.** Here we use matrix factorization techniques common to recommendation systems in order to explore respondents' interpretation of the original survey questions by plotting each item according to its position on the two most distinguishing latent dimensions. This enables us to visually and qualitatively analyze image characteristics that represent axes of difference in respondent interpretations of each context. We demonstrate this with data from R100, which best predicts both individual and aggregate comparative assessments. Figures 7, 8 and 9 plot the items and their corresponding rated quartiles along the two most descriptive latent dimensions for each context.

From visual inspection of safety perceptions for Google Streetviews, we note that the amount of disarray constitutes the most informative dimension, with more chaos being associated with lower perceived safety (Molnar et al. 2004). It seems as if when asked to rate image safety, respondents simulated scenarios where those images manifesting greater neglect and disorganization offered more perceived opportunities for unexpected and unwanted interactions. The second most informative dimension is the amount of pavement or greenery in the Streetviews, with more pavement associated with lower safety. This plot reveals respondent interpretations of safety in Streetview images by demonstrating that most interpreted safety as from potential dangerous human or animal agents, hidden amongst the disarray, but others from cars that rule the pavement. We note that these dimensions are modestly correlated, leading to a negatively sloped pattern.

By graphically plotting masterpiece likeability, we note that the most informative dimension regarding whether respondents like a work is whether it renders a scene from a human environment. The second most important dimension is the realism or abstraction of its imagery. When asked if they liked a masterpiece, respondents implicitly asked themselves whether the painting was human-scale or realistic or both. If so, they tended to like it. These two dimensions are not highly correlated, as evidenced by their cloud-like distribution.

When we examine the plots of funny animal GIFs, we see that the most informative dimension is whether animals are in a human environment or exhibiting human-like behavior, jogging on the treadmill or enjoying a cup of morning joe, with animals mimicking humans rated funnier. The second most informative dimension is whether the animal is situated in a domestic or wild and natural setting, with domesticated animals systematically perceived as more funny. These two dimensions of interpretation are highly correlated among our GIFs, which explains why the images lie along a loose, straight line. Even though these two dimensions help us to interpret respondent meanings when they rated an animal GIF as funny or not, only the first would be required to distinguish them from one another in terms of humor or hilarity.

______________________________

Figure 7 about here

______________________________

______________________________

Figure 8 about here

______________________________

______________________________

Figure 9 about here

______________________________

**Discussion**

In this paper we introduced and critically evaluated simple, sparse surveys from the world of recommendation systems for belief, opinion and preference measurement in the social sciences. Recommendation systems deploy simple surveys to their users in order to improve personalized recommendations and global services. Here we showed how matrix factorization can estimate the inherent structure in partial and incomplete survey responses to reveal latent structures through a relatively small number of simple questions. We used a crowd sourced Amazon Mechanical Turk survey to measure the performance of recommendation system models in a variety of contexts and then estimated the relative performance of binary choice, Likert-scale, continuous slider, and pairwise ranking for predicting unasked comparative assessments.

Across all contexts, we found evidence that partial surveys convey substantial information for measuring and predicting subjective assessments and preferences. In comparison with traditional surveys, which must typically be completed to merit inclusion for analysis, follow a balanced factorial design (Rossi and Nock 1982) or statistically interpolate missing values (Little and Rubin 2014), simple surveys are constructed to be performed rapidly, incompletely and cheaply. They nevertheless reveal substantial information that can predict previously unasked individual and aggregate responses. Our analysis revealed that different surveys exhibit distinctive properties. Ratings on a continuous scale take approximately 50% longer to complete than other simple surveys as they appear to require more memory and self-calibration to answer consistently. But when consistency and sensitivity is critical, continuous scale questions outperform other simple survey types in anticipating both individual and aggregate assessments. The distribution of ratings within respondents suggests that continuous scales allow them to reserve scale space to report surprisingly good or bad impressions as they face new, unexpected items. Even though variation in response to continuous scales correlates strongly with five-point scales, continuous scales retain substantially more signal about individual and collective assessments than five-point scales.

Pairwise comparisons and binary choice questions should theoretically convey the same amount of information—1 bit per question—and they took approximately the same amount of time to fill out. Nevertheless, they performed very differently in anticipating individual versus aggregate comparative assessments. Pairwise comparisons performed very well for predicting individual comparisons, but consistently the worst at predicting aggregate comparisons. A number of recent research designs have proposed pairwise comparisons as an effective method for identifying aggregate comparisons, such as Hidalgo's *Streetscore* framework for eliciting stable collective assessment of streetview images (Salesses, Schechtner, and Hidalgo 2013; Naik et al. 2014)[5]. Pairwise comparisons have also been used to crowdsource and crowdrank good ideas for collective adoption, as in Salganik's wiki-survey framework (Salganik and Levy 2015; Salganik 2017).[6] Binary choice surveys, in contrast, systematically performed worst at predicting individual comparisons, but compete with continuous scales as best at predicting aggregate comparisons. Netflix recently adopted binary choices (thumbs up/down) to predict individual assessments. Our study suggests that Netflix customer would better benefit from pairwise comparisons, and crowdsourcing good ideas or aggregate responses might better employ binary choice designs.

Our matrix factorization approach also shed light on latent dimensions that could allow analysts to quickly decipher respondent interpretation of questions posed. With the visual items we employed in these simple surveys, plotting them on the first and second most discriminating dimensions immediately suggested respondent interpretations. For example, in responding to questions about "safety", respondents appear to have considered the concept in at least two different ways—for themselves (e.g., crime associated with disarray) and for their children (e.g., traffic accidents associated with cars on pavement). In responding to classic paintings, respondents revealed that they tend to like realistic, human-scale artworks, although these preferences account for less of the overall variation in respondent ratings than safety assessments. We also found that domesticated and human-like activity best characterized the gradient of responses regarding whether animals GIFs were perceived as funny.

Simple and incomplete surveys have obvious limitations in comparison with complete rating and ranking surveys, especially when performing statistical inference on respondent positions over modest sets of items. Unbalanced surveys make inference more difficult. Nevertheless, matrix factorization makes sparse imputation possible that is inconceivable with traditional approaches. As a result, we reduced our test error below 15% for predicting individual comparisons and 20% for predicting aggregate comparisons where many of the compared items had never been ranked previously by the individual. This highlights how a critical aspect of simple survey design is that questions be distributed across respondents enabling matrix factorizations to identify similar respondents and similar items. Here we did this by randomizing questions posed to each respondent. In the case of simple surveys deployed in the wild or on the web, where most respondents provide different numbers of responses, posing questions nonrandomly in order to maximize the likelihood that factorization can infer answers to questions not asked will be an important target for future research.

We argue that simple surveys can add substantial insight when the number of issues to be collectively addressed is impractical or impossible with established survey designs. Consider the domain of political opinions, where complex configurations of political preferences and associations might require many more items than could be elicited by an exhaustive survey. Surveys that demand complete responses to a modest number of questions enable statistically sound inference, but simple surveys could reveal a more nuanced arrangement of contingent understandings that could chart novel pathways to possible political agreement and coalition-building.

The potential for simple, partial surveys to reveal a wide range of subjective beliefs, preferences, opinions and ideas suggest new frontiers for survey interactivity, such as inviting respondents to add new items in ongoing, evolving 'wiki-surveys' (Salganik and Levy 2015). Additional efficiencies could be achieved by harnessing approaches from the computer science subfield of active learning (Castro et al. 2009) and computer adaptive testing approaches (Wainer et al. 2000) to create intelligent surveys, where answers to previous questions are used to optimally select queries sequentially in order to gather the most information with the fewest elicitations. Simple, partial surveys create the design flexibility that could allow rethinking sampling design as algorithm design, optimizing samples to confirm or refute specific models of the social world (Suchow and Griffiths 2016) . We encourage additional exploration with simple surveys, which could be 'gamified' for use in settings well beyond the reach of exhaustive survey instruments (Hamari, Koivisto, and Sarsa 2014) where it may be important to infer less information about more people than more about less.

# Appendix

We assume that our rating matrices are sparse, or have many unobserved entries. We denote observed entries as $X_{obs}$ and the indices of those entries as $(i,j) \in \Omega \subset \{1, \cdots n\} \times \{1, \cdots, p\}$. In order to impute unobserved entries, we use the *R* package `SoftImpute`, which minimizes a simple quadratic loss function to estimate factors *U* and *V*:

$$\min_{U,V} \sum_{(i,j) \in \Omega} (X_{ij} - u_i v_j)^2 + \gamma \left( U_F^2 + V_F^2 \right) \quad (1)$$

where the first term in the equation is the data fitting term and the second is the regularization penalty to prevent overfitting. Further, $u_i \in \mathcal{R}^{1 \times k}$ denotes the $i_{th}$ row of *U*, $v_j \in \mathcal{R}^k$ denotes the $j_{th}$ row of *V*, and $u_i v_j = (UV)_{ij} \in R$ denotes the inner product between the two. Finally, $\cdot_F$ indicates the Frobenius norm of a matrix, or the square root of the sum of absolute squares of its elements (Golub and Van Loan 1996, 55), and *γ* ≥ 0 controls the relative weight given to regularization. Using this quadratic loss function, greater regularization, *γ* > 0, smooths the *UV* matrix solution, which tends to prevent overfitting by reducing the construction of outliers.

The minimization problem above has two parameters, rank *k* and regularization penalty weight *γ*, which need to be set. If we ignore regularization (*γ*=0) and set a small *k* (*k* ≪ min{*n,m*}), we have a pure 'low rank' approximation because *k* is low. Alternatively, if we increase regularization (*γ* > 0) and set a *k* that equal to or larger than our data (*k* ≥ {*n,m*}) then this reduces to a pure 'low norm' approximation because the regularizing matrix norms are forced to be low through optimization / minimization. In practice the parameters are typically selected using out-of-sample evaluation or cross-validation. Out-of-sample methods randomly sample from the observed values (the training set) and then evaluate the performance of a given model (i.e. a model with a fixed value of $k$ and/or $\gamma$) on the remaining observations or validation set. With cross-validation, this is done repeatedly, with distinct, rotating bins of training and validating cases. The idea here is to simulate the performance of the model on out-of-sample data. Subsamples may be randomized over rows, columns or indices.

It is also useful to keep aside data that is part of neither training nor validation set. This is typically referred to as the testing data, although definitions of testing and validation sets are sometimes reversed. This testing set is assumed to represent the "ground truth" and predictions from different algorithms are compared by measuring error in predicting the data in test set or 'test error' (see Lee and Seung 1999; Koren, Bell, and Volinsky 2009 for a thorough discussion of these techniques).

## Author Bios

**James A. Evans (Corresponding Author)**
James Evans is Professor of Sociology, Director of Knowledge Lab, and Faculty Director of Computational Social Science at the University of Chicago, and an external Professor at the Computation Institute. His research uses large-scale data, machine learning and generative models to understand how individuals and collectives think and what they know, with a special emphasis on science, technology and innovation. Evans designs observatories for understanding that fuse data from text, images and other sensors with results from interactive crowdsourcing and online experiments. Evans received his PhD in Sociology at Stanford University in 2004.

Email: jevans@uchicago.edu

**Nandana Sengupta**
Nandana Sengupta is Assistant Professor in the School of Policy and Governance at Azim Premji University. Nandana's research revolves around improving the predictive performance of traditional econometric models using modern statistics and machine learning algorithms. She's primarily interested in using these techniques in public policy applications.

Nandana received a PhD in Economics at the Tepper School of Business at Carnegie Mellon University in 2015. She also holds an MSc. degree in Development Economics from IGIDR, Mumbai and a BSc. degree in Physics from St. Stephen's College, Delhi. Prior to joining Azim Premji University, Nandana was a postdoctoral scholar at the Knowledge Lab at University of Chicago's Computation Institute.

Email: nandana.sengupta@apu.edu.in

**Nathan Srebro**
Nathan Srebro is Professor at the Toyota Technological Institution of Chicago, and part time faculty in Computer Science and the Committee on Computational and Applied Mathematics at the University of Chicago. Srebro is interested in statistical and computational aspects of machine learning and their interaction. He has done theoretical work in statistical learning theory and in algorithms, devised novel learning models and optimization techniques, and has worked on applications in computational biology, text analysis, collaborative filtering and social science. Srebro obtained his PhD from the Massachusetts Institute of Technology in 2004.

Email: nati@ttic.edu

# Citations

**Endnotes**

1. Other simple survey designs exist, which we were unable to evaluate in this paper, including best-worst scaling pioneered by (Louviere, Flynn, and Marley 2015) where respondents specify best *and* worst responses.
2. We do not use full ranking surveys because they are not "simple" in that they require sustained focus to normalize responses regarding a population of numerous items.
3. Missing values estimation using matrix factorization performs far superior to existing social science methods for imputing missing data under conditions of very sparse matrix reconstruction with many missing values (Nandana Sengupta, Madeleine Udell, Nathan Srebro, James Evans, n.d.).
4. Even though the same number of rating and response questions are asked (80) to predict the final 20 comparisons elicited afterwards, note that distinct information is obtained from ratings and rankings. If ratings are issued consistently, this would provide accurate information about the 3160 comparisons between them, compared with only 80 explicit comparisons asked with the pairwise comparison method. If ratings are inconsistent applied, however, then 80 explicit pairwise comparisons may provide more effective prediction.
5. This is implemented online in Hidalgo's Place Pulse comparison survey engine at http://pulse.medial.mit.edu
6. This is implemented online in Salganik's All Our Ideas online wiki-survey engine at http://www.allourideas.org

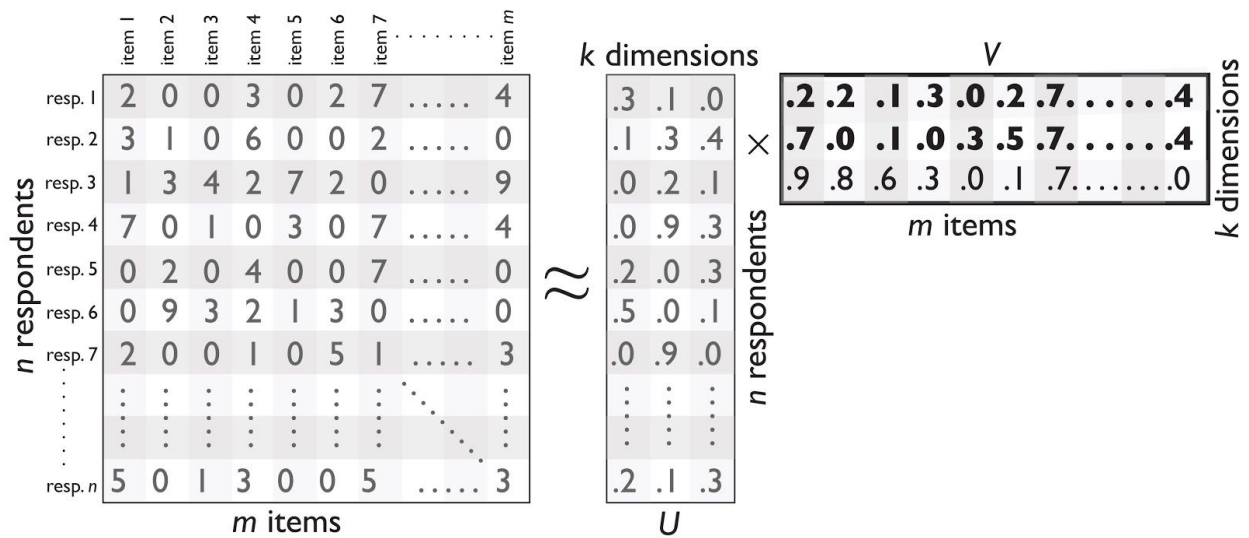

**Figure 1. The problem for matrix completion.** Schematic figure that illustrates the structure of the problem matrix completion algorithms seek to solve—how to project respondents and items into a k-dimensional space that predicts answers to unasked questions. The solution is $UV$, where $U$ is an *n*-by-*k* matrix of values and $V$ is a *k*-by-*m* matrix of values, shown here for the example $k = 3$. The values of $U$ can be understood as coordinates for each person in a $k$ dimensional latent space; and the values of $V$ as the coordinates for each survey item in that same space. We bold the first two dimension of $V$, which we use to plot each item in Figure 7. In practice, the matrix would typically be much sparser, containing 90 to 99.999% zeros.

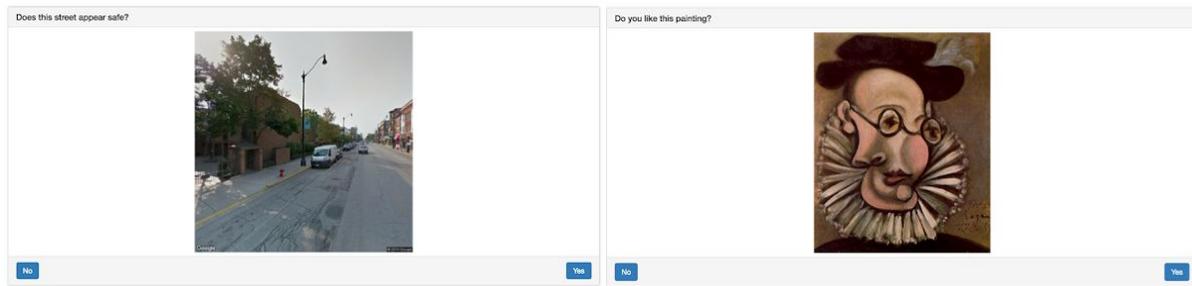

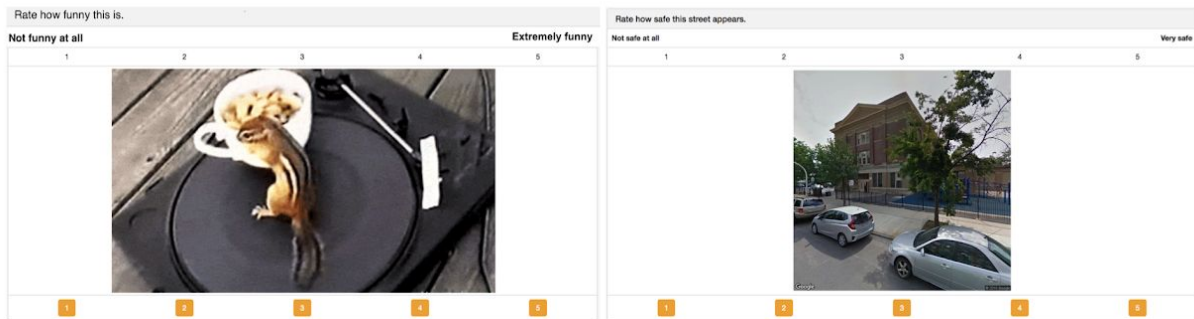

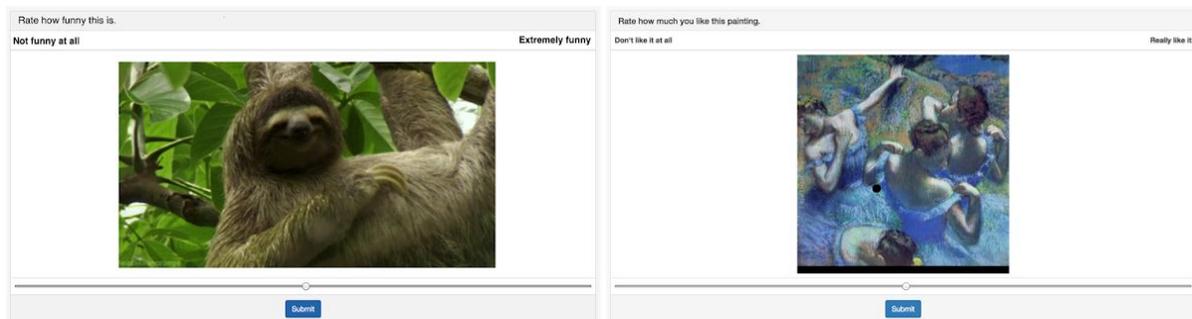

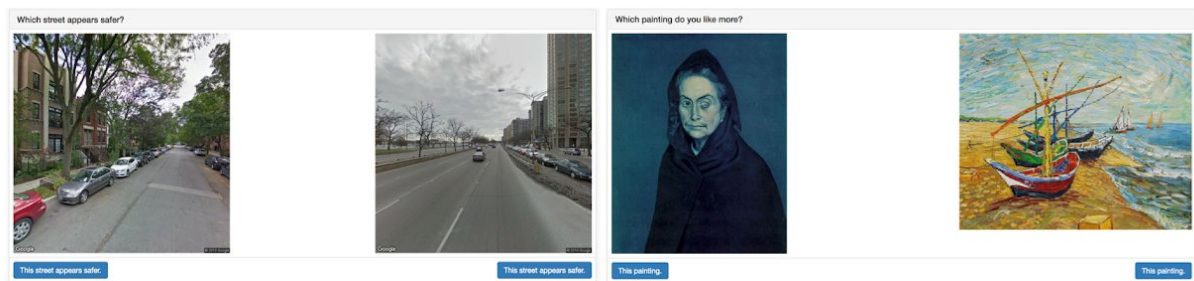

**Figure 2. User interface for the simple surveys.** Examples of the queries MTurk respondents were presented in the three different contexts for the four survey types.

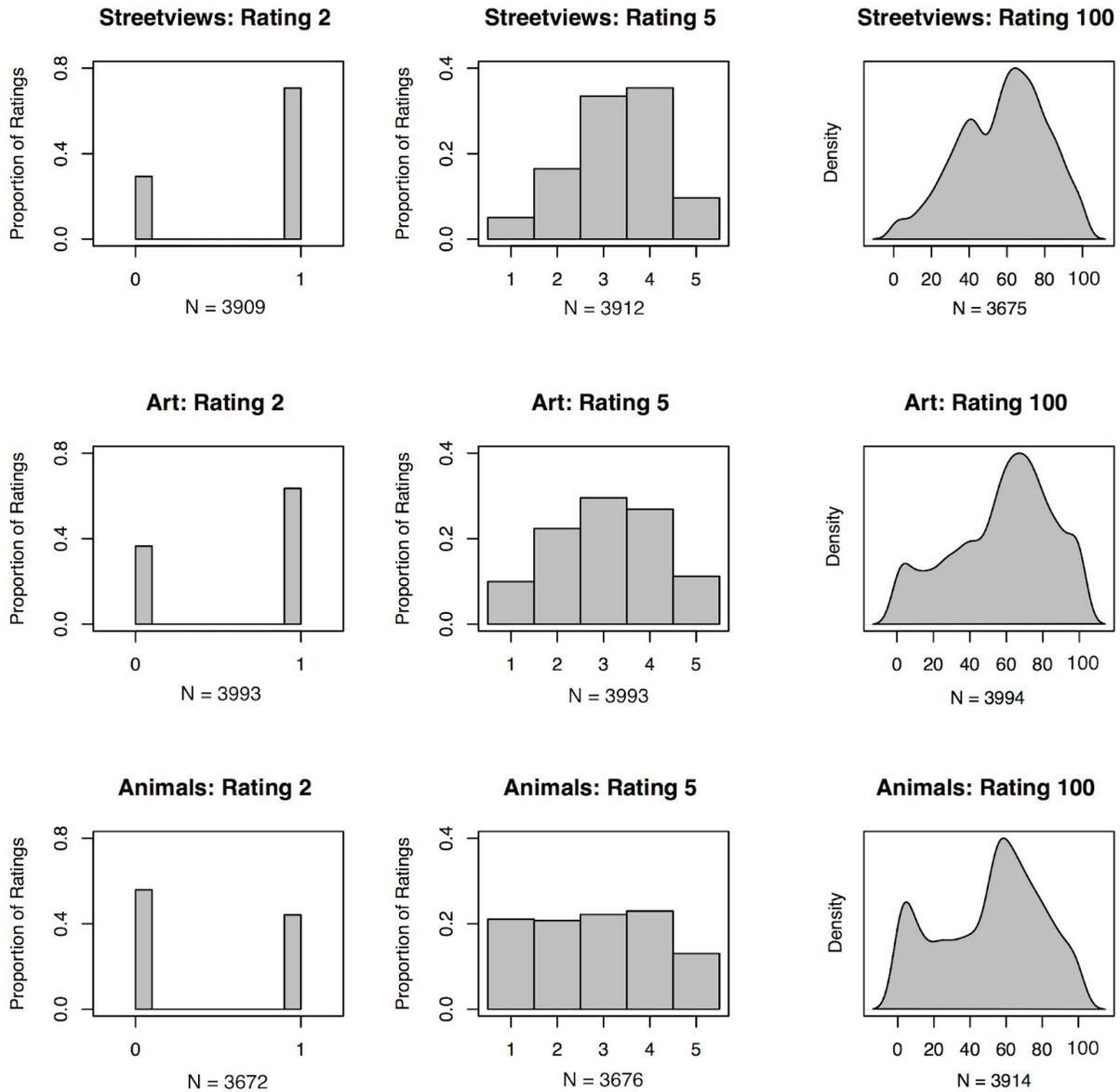

**Figure 3. Distribution of Ratings. Histograms and** density plots for all ratings data collected in all three contexts. The total number of ratings collected for each ratings task is noted at the bottom of each subplot. Rating 2 refers to binary choice questions , for example, "Does this street [Streetview panorama] appear safe?" ["Yes" or "No"]. Rating 5 refers to a 5 point Likert scale for example, "Rate how much you like this painting [masterpiece image]". ["Don't like it at all" – 1 – to "Really like it" – 5]. Finally, we elicit preferences with ratings on a continuous "slider" scale, for example, "Rate how funny this [animal GIF] is". [Slider calibrated between "Not funny at all" to "Extremely funny"]. While the "continuous" scale appears as a continuous slider from the respondents perspective, it is rastered to a discretized scale from 1 to 100, which is what the the label of Rating 100 reflects.

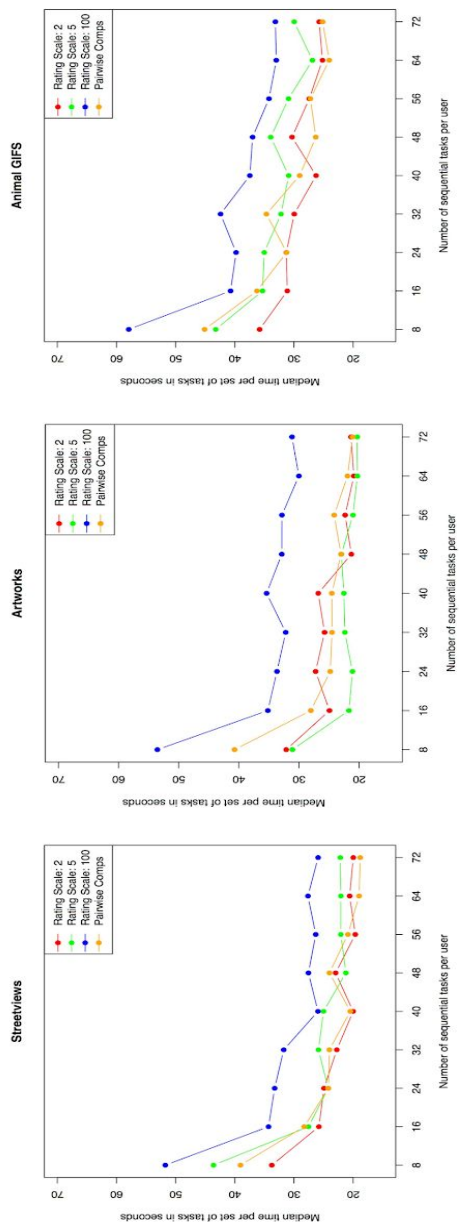

**Figure 4. Median time for every set of next 8 queries in seconds.** Median time to complete each survey, where respondents are incentivized to do more in less time, stands as a rough proxy of the cognitive load required to perform each of the four different query types. Medians are compared because they are less affected by extreme values. As expected, median times for all tasks across all contexts decrease as respondents answer more questions. R100 takes consistently more time than all other query types, because choosing a single point on a continuous slider intuitively takes respondents longer to calibrate and recall.

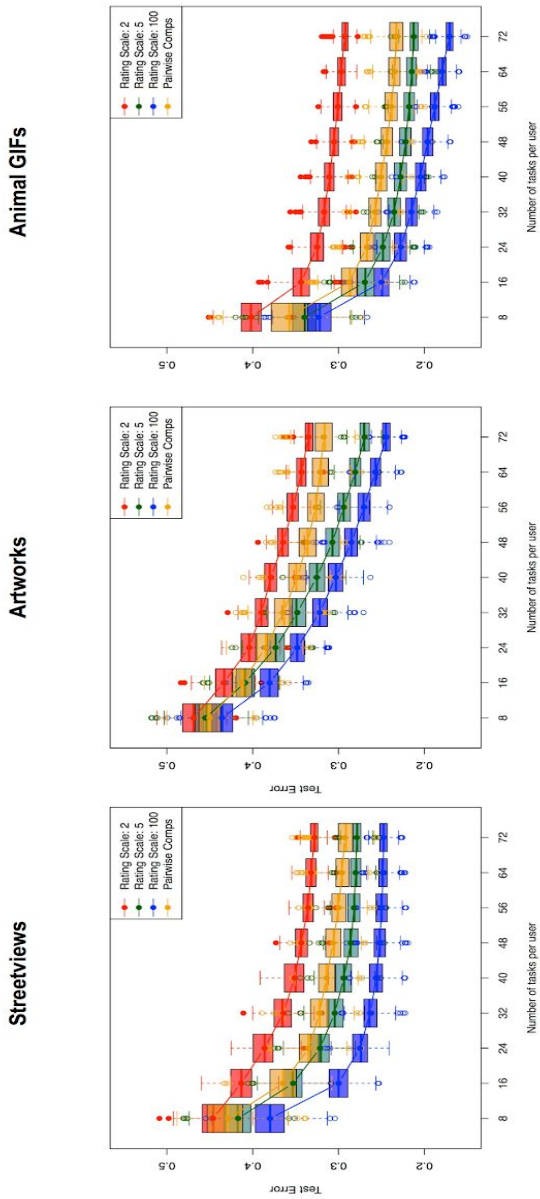

**Figure 5. Predicting Individual Preferences.**
Prediction estimates for each respondents' direct comparisons using a matrix completion model built on the number of queries listed on the *x*-axis. Note that when very few items are considered, such as 8, these techniques still beat a random coin toss and sometimes perform much better. This implies that even with very sparse data from each respondent, consolidating data from a moderate number of respondents can be used to make meaningful predictions.

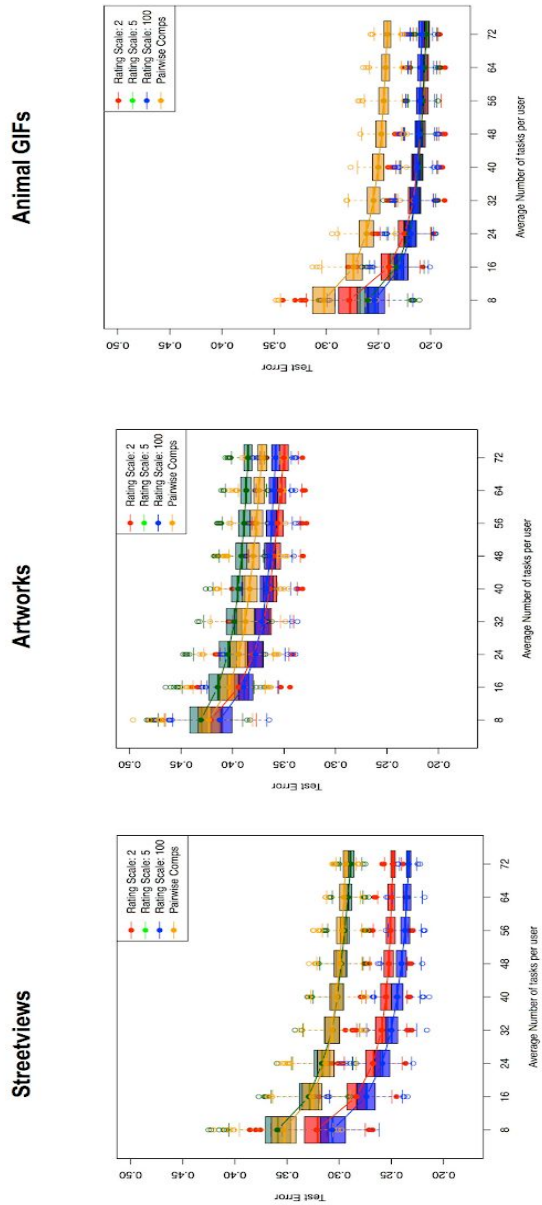

**Figure 6. Predicting Global Rankings.**
Prediction estimates for global rankings using a matrix completion model built on the number of queries listed on the *x*-axis. Note that variations in predicting global rankings versus individual preferences in suggest that some contexts are less suited to aggregation. In contrast to the context of visual perceptions of safety, individual preference predictions are consistently better than global rankings in the contexts of likability of masterpieces and the funniness of animal GIFs. Also, R2 which performed worst in predicting individual preferences, competes with R100 for best predicting global rankings.

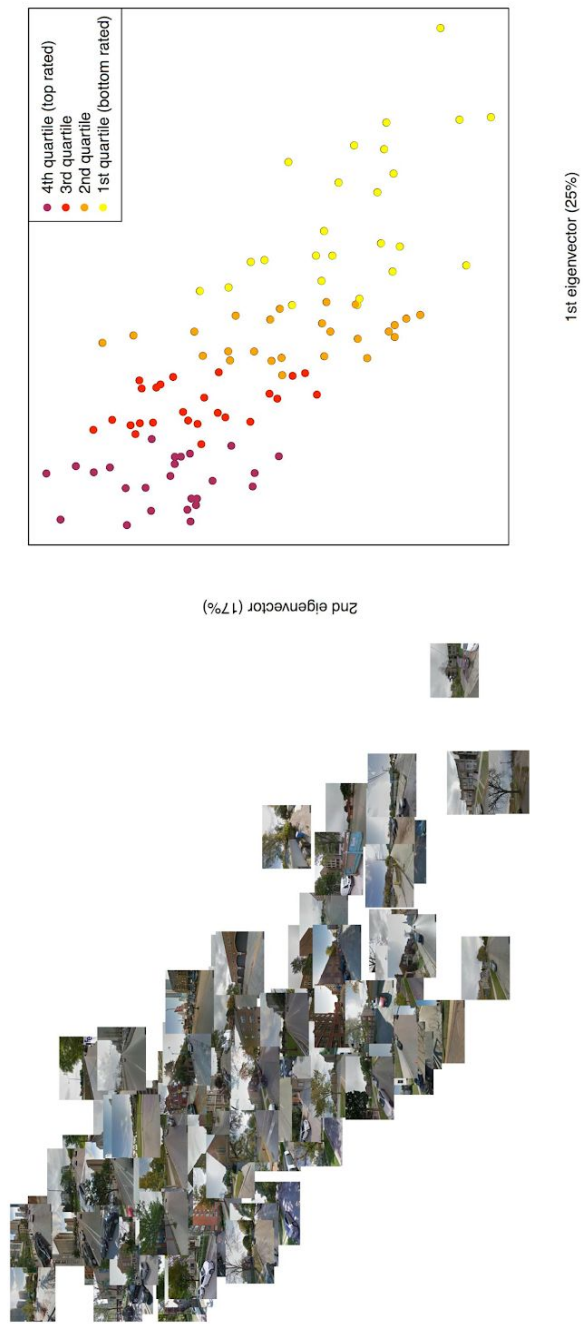

**Figure 7. Latent Dimensions Representation of Streetviews.** Streetviews embedded on to the two latent dimensions along which R100 varies most. The two latent dimensions explain 41% of total variation in ratings across all respondents. The embeddings help us to visually analyze image characteristics that represent axes of difference in respondent interpretations of visual perceptions of safety.

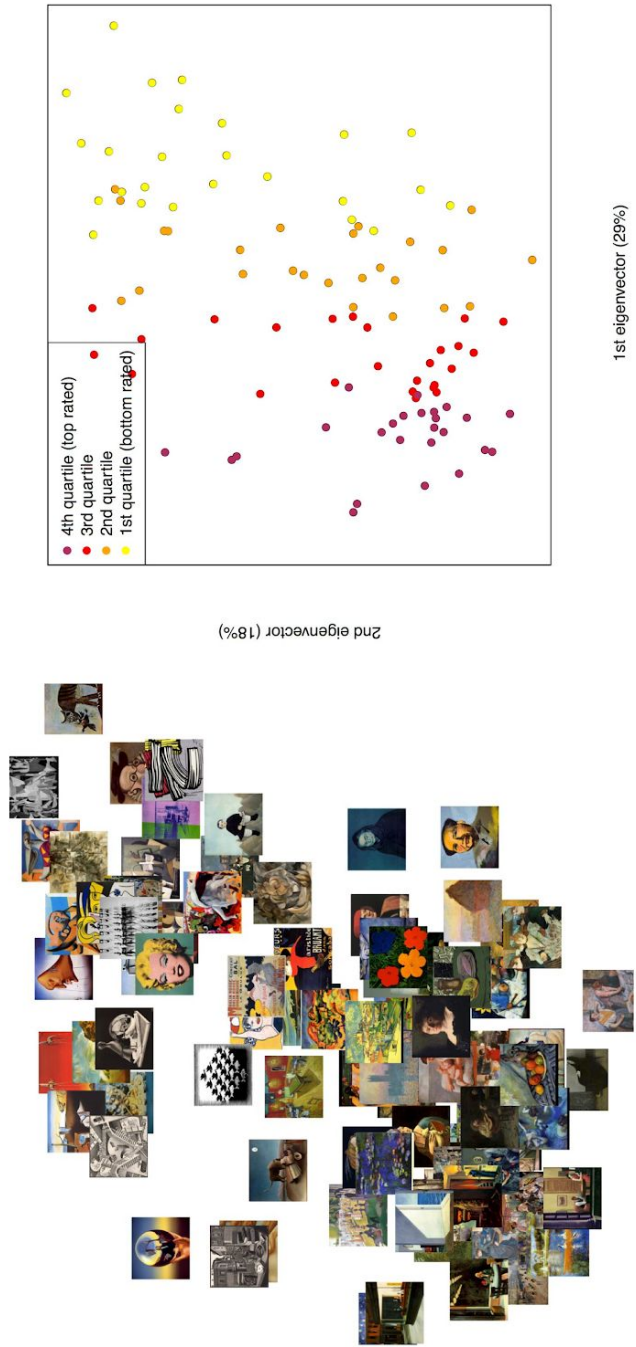

**Figure 8. Latent Dimensions Representation of Artwork.** Artwork embedded on to the two latent dimensions along which R100 varies most. The two latent dimensions presented explain 47% of total variation in ratings across all respondents. The embeddings help us visually analyze image characteristics that represent axes of difference in respondent interpretations of artwork likeability.

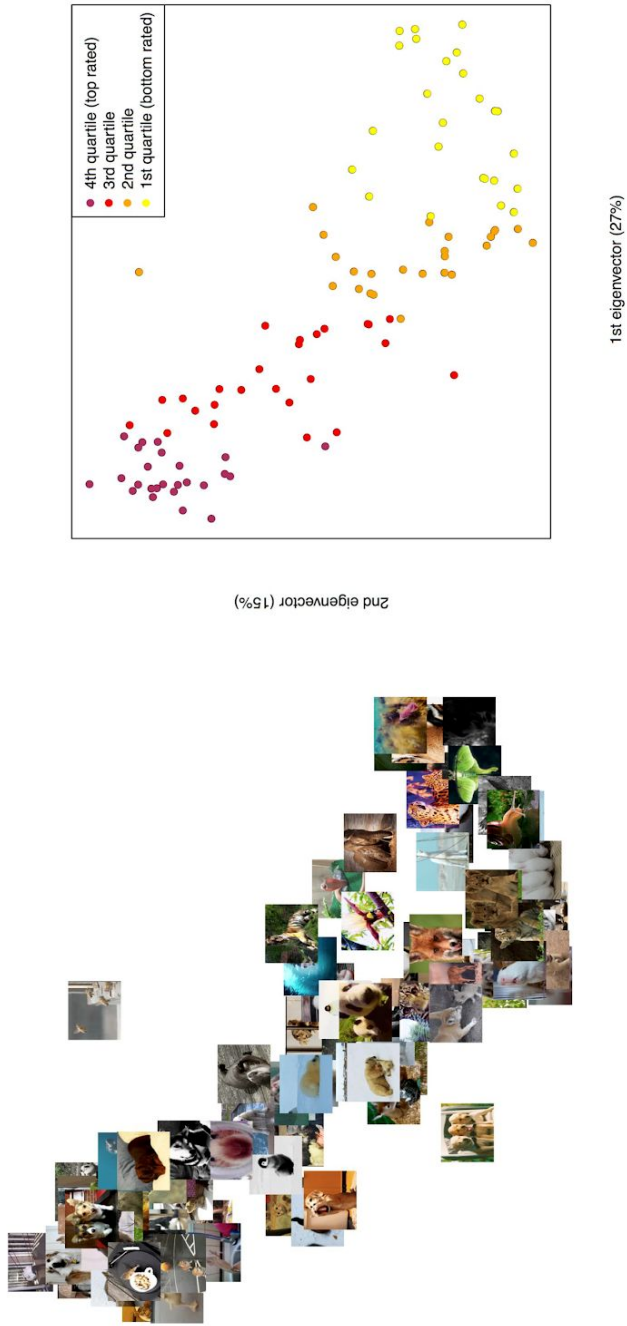

**Figure 9. Latent Dimensions Representation of Animal GIFs.** Animal GIFs embedded on to the two latent dimensions along which R100 varies most. The two latent dimensions presented explain 42% of total variation in ratings across all respondents. The embeddings help us visually analyze image characteristics that represent axes of difference in respondent interpretations of Animal GIFs hilarity.